\documentstyle[aps,epsf,floats,twocolumn]{revtex}
\begin{document}
\draft
\title{ Diffusion and localization in chaotic billiards}
\author{Fausto Borgonovi$^{[a,c,d]}$, Giulio Casati$^{[b,c,e]}$ 
Baowen Li$^{[f,g]}$}
\address{
$^{[a]}$ Dipartimento di Matematica, Universit\`a Cattolica, 
via Trieste 17, 25121 Brescia, Italy  \\
$^{[b]}$  Universit\`a di Milano, sede di Como, Via Lucini 3, Como, Italy\\
$^{[c]}$ Istituto Nazionale di Fisica della Materia, Unit\`a di Milano,
via Celoria 16, 22100 Milano, Italy \\
Istituto Nazionale di Fisica Nucleare, Sezione di Pavia$^{[d]}$,
Sezione di Milano$^{[e]}$,\\
$^{[f]}$ Department of Physics and Centre for Nonlinear and Complex
Systems, Hong Kong Baptist University, Hong Kong  \\
$^{[g]}$ Center for Applied Mathematics and Theoretical Physics,
University of Maribor, Krekova 2, 2000 Maribor, Slovenia
  \medskip\\
\parbox{14cm}{\rm
We study analytically and numerically the classical diffusive process 
which takes place in a chaotic billiard. This allows to 
estimate the conditions under which the statistical 
properties of eigenvalues and eigenfunctions can be 
described by Random Matrix Theory. In particular the 
phenomenon of quantum dynamical localization should be 
observable in real experiments.
}}
\maketitle
\narrowtext

 One of the main modifications that quantum mechanics introduces in
our classical picture of deterministic chaos is ``quantum dynamical
localization'' which results e.g. in the suppression  of
chaotic diffusive-like process which may take place in systems under
external periodic perturbations. This phenomenon,  first
pointed out in the model of quantum kicked rotator \cite{CCFI}, is now
firmly established and observed in several laboratory 
experiments \cite{Salcazzo1}.

    For conservative Hamiltonian systems the question of localization
is much less investigated. 
The situation here is much more intriguing : from one hand, in 
a conservative system, one may argue that there is always localization
due to the finite number of 
unperturbed basis states effectively coupled 
by the perturbation; on the other hand a large amount of
numerical evidence indicates that quantization of classically 
chaotic systems leads to results 
which appear in agreement with the predictions
of Random Matrix Theory (RMT) \cite{Bohigas}.

   Recently the problem of localization in conservative systems has been
explicitely investigated. In particular, on the base of Wigner band
random matrix model, conditions for localization were explicitely
given together with the relation between
localization and level spacing distribution \cite{Salcazzo2}.

Billiards are very important models in the study of 
conservative dynamical systems
since they provide clear mathematical examples of classical chaos 
and their quantum properties have been extensively studied theoretically 
and experimentally. Moreover they are becoming increasingly relevant
for the study of optical processes in microcavities which may lead 
to possible applications such as the design of novel microlasers
or other optical devices \cite{DS}.

In this paper we focus our attention on a two dimensional chaotic billiard:
 the 
Bunimovich stadium, and study the classical diffusive process which 
takes place in angular momentum. This will allow us to predict the 
conditions for quantum localization and therefore the conditions 
under which the standard Random Matrix Theory is not applicable.

 We consider the motion of a  particle 
having mass $m$, velocity $\vec v$ and elastically bouncing
inside  the stadium shown in Fig 1. We denote with $R$ the radius of 
the semicircles and with $2 a$ the length of the straight segments.
The total energy is $E = m\vec{v}^2/2$.

The statistical properties of the billiard are
controlled by the dimensionless parameter $\epsilon=a/R$ and, for
any $\epsilon>0$, the motion is ergodic,  mixing and 
 exponentially unstable with Lyapunov exponent  $\Lambda$ which, for
small $\epsilon$, is given by
\cite{Benettin}
$\Lambda \sim \epsilon^{1/2}$.

For the analysis of classical dynamics, a typical choice of canonical 
variables 
is $(s, v_t)$ where $s$
measures the position  
along the boundary 
of the collision point
and $v_t$ is the tangent velocity.
These variables however, are quite difficult to treat from the quantum
point of view. For this purpose it is convenient instead 
to consider 
$l$, the angular momentum
calculated with respect to the center of the
stadium, and the angle $\theta$ which describes, together with
$r(\theta)$, the position of the particle in the usual polar
coordinates.
It is important to stress that with this choice of 
variables, the invariant measure
$d\mu = ds dv_t$ is preserved only to order $\epsilon$,
that is $d\mu = ds dv_t = d\theta d l + o(\epsilon)$

At a given energy $E$, the angular momentum 
 must satisfy 
the relation
$\vert l\vert < l_{max} =  (R+a)\sqrt{ 2Em}$. It is therefore
convenient to introduce the rescaled quantity $L=l/l_{max}$.
Then the classical motion takes place on the cylinder
$0 \leq \theta<2\pi$,  $-1<L<1$.

It is expected that for $\epsilon<<1$ a diffusive process will
take place in angular momentum with a diffusion coefficient 
$D=D(\epsilon)$. In order to obtain an estimate 
for $D(\epsilon)$ we now derive 
an explicit expression for the boundary map 
in 
the variables $(L,\theta)$.
The change $\Delta L$ after a  collision with the
boundary  
can be easily obtained 
to order $\epsilon$, by 
neglecting collisions with straight lines 
and 
by taking into account that in the collision
only the normal velocity $ v_n = \vec{v} \cdot \vec{n}$ changes
the sign. Here $\vec{n} \simeq \vec{e}_r +\epsilon\sin\theta
sign(\cos\theta) \vec{e}_{\theta}$, $\vec{e}_r$ and $\vec{e}_\theta$
being the usual polar coordinates unit vectors.
One then get :

\begin{equation}
\Delta L = \bar{L}- L = -2\epsilon \sin\theta sign(\cos\theta) sign(L)
\sqrt{1-L^2}
\label{deltal}
\end{equation}

On the other side the change in $\theta$,
to zero order, is given by :

\begin{equation}
\Delta \theta = \bar{\theta} - \theta = \pi -2\arcsin ( {\bar{L}})
\label{deltat}
\end{equation}

According to a standard procedure \cite{LL} we introduce a
generating function  $G(\bar{L},\theta)$ in such a way that
the map defined by

\begin{equation}
L ={ {\partial G}\over {\partial \theta}} ; \qquad
\bar{\theta} = { {\partial G}\over {\partial {\bar L}}}
\label{eql}
\end{equation}

coincides with $\Delta L$  at first order in $\epsilon$ and
with $\Delta \theta$ at zero-th order.
The generating function is given by:

\begin{equation}
G(\bar L,\theta) = (\theta+\pi){\bar L} -
2\int^{\bar{L}}  dL \arcsin L 
+\epsilon
g(\bar L) \vert \cos \theta \vert 
\label{gf}
\end{equation}

where
$g(\bar L) = 2 sign (\bar L) (1-{\bar {L}}^2)^{1/2}.$
The generated (implicit) area--preserving map is

\begin{equation}
\begin{array}{l}
\bar{L} = 
L -2\epsilon \sin\theta sign(\cos\theta)sign(\bar L)
( 1-{\bar L}^2 )^{1/2} \\
\bar{\theta} = \theta + \pi -2 \arcsin(\bar L) + \epsilon g'(\bar L)
\vert \cos \theta \vert 
\label{mapc}
\end{array}
\end{equation}

 By taking  the local approximation 
in the angular momentum, the map (\ref{mapc}) writes :

\begin{equation}
\begin{array}{l}
\bar{L} = 
L -2\epsilon \sin\theta sign(\cos\theta)sign(\bar L)
\sqrt{1- L_0^2}\\
\bar{\theta} = \theta + \pi -2 \arcsin(\bar L) 
\label{mapl}
\end{array}
\end{equation}

which remains area--preserving and can be easily iterated
(here $L_0$ is the initial angular momentum).

The agreement of map (\ref{mapl}) 
with the true dynamics can be numerically checked and it is shown
in Fig.2 where we plot 
$L^* = (\bar{L}-L)/(2\epsilon\sqrt{1-L_0^2})$ against 
$\theta$. Points represent billiard dynamics while the full line 
is the function
$f(\theta) = - \sin\theta sign(\cos\theta)$.

Notice that the function $f(\theta)$ 
is periodic  of period $\pi$
and has a discontinuity 
at $\theta = \pi/2$.
This  gives to the map
(\ref{mapl})
a structure very close to the sawtooth map which is known\cite{DMP}
to be chaotic and diffusive
with a diffusion rate $D$ which, 
for small values of the 
kick strength
$\epsilon$, is given by $D\sim \epsilon^{5/2}$.
This behaviour, according also to our numerical computations,
 appears
to be generic for maps which have such type of      
discontinuity.

We may proceed now to a 
numerical investigation of the diffusive process.
To this end 
 we consider a distribution of particles
with given  initial
$L_0$ and random phases $\theta$ in the interval $(0,2\pi)$ and integrate
the classical equations of motion inside the billiard.
In Figs. 3a,b we present the behaviour of
$\Delta L^2  = \langle L^2 \rangle  - \langle L_0 \rangle ^2$
as a function of the number of collisions  $n$
 and the distribution function $f_n (L)$ at
fixed $n$ 
as a function of $(L - L_0)$.
As it is seen, $\Delta L^2$  grows 
diffusively 
and the distribution function is in good agreement
with a Gaussian\cite{nota}.
In particular
the dependence $D=D(\epsilon)$ 
of the diffusion coefficient 
can be easily computed and the result
$D=D_0 \epsilon^{5/2}$ (see Fig.4)
is in agreement with predictions 
of  map (\ref{mapl}) with $D_0 = 1.5$.

The analysis of the classical diffusive process allows to make 
some predictions concerning the quantum motion and in particular
to estimate the conditions under which 
the quantum localization 
phenomenon will take place \cite{chirikov}.
First of all, in order that any quantum diffusive process may 
start 
it is necessary to be above the perturbative regime. In particular
the level number must be sufficiently high so that the
De Broglie wavenumber $k$ of the corresponding 
wavefunction must satisfy the relation $k > 1/a$.
This implies 
$E > E_p = \hbar^2 /2ma^2 $ which is the energy necessary to confine 
a quantum particle inside a box of length $a$. 
Using the  well known  Weyl formula 
for the  total number 
of states with energy less than $E$ \cite{Bohigas} 

\begin{equation}
\langle N(E) \rangle \approx { {m\cal{A}}\over {2\pi\hbar^2}} E 
\approx {1\over 8} m \left( {R\over \hbar} \right)^2 
E 
\label{tf}
\end{equation}

where ${\cal A}$ is the area of a quarter of billiard,
and keeping only the leading term, 
we obtain that in order to be in a non perturbative regime 
we have to  consider  level numbers  

\begin{equation}
N \gg N_p \simeq { {1}\over {16\epsilon^2}} 
\label{SB}
\end{equation}

We call $N_p$ perturbative border.

According to the well known arguments \cite{CIS},
above the perturbative border (\ref{SB}) quantum diffusion 
in angular momentum takes place
with a diffusion coefficient close to the classical one.
This diffusion proceeds 
up to a time 
$\tau_B \sim D_{eff}/\hbar^2 $
after which diffusion will be suppressed  by
quantum interference. 
This time is related to the uncertainty principle. Namely,
for times less than $\tau_B$ the discrete spectrum is not
resolved and the quantum motion mimics the classical 
diffusive motion \cite{CIS,CC}.
Here 
$D_{eff} = D_0 \epsilon^{5/2} 2mER^2$ 
is the 
classical diffusion coefficient in real (not scaled) angular
momentum.

The nature of the quantum steady state will depend crucially 
on the ergodicity parameter \cite{CC}

\begin{equation}
\lambda^2 = { {\tau_B}\over{\tau_E}}
\end{equation}

where 
$\tau_E = l^2_{max}/D_{eff} \simeq 2mER^2/D_{eff}$
is the  ergodic relaxation time.

For $\lambda \ll 1$ the quantum steady state is localized 
while for $\lambda \gg 1$ we have quantum ergodicity.
The critical value $\lambda = 1$ leads to 
$l_{max }\hbar = D_{eff}$ that is 
$E = E_{erg} = \epsilon^{-5} D_0^{-2} \hbar^2/2mR^2$. 
We then have :

\begin{equation} 
 N =  N_{erg} \simeq  {{1}\over {16 D_0^2 \epsilon^5 }} 
\label{erg}
\end{equation}

It follows that only for $N > N_{erg}$ 
there is quantum ergodicity and therefore one   
expects statistical properties
of eigenvalues and eigenfunctions to be described by RMT. 
Instead for $N < N_{erg}$, even if $N \gg N_p$, namely  very 
deep in quasiclassical regions, statistical properties will 
depend on parameter 
$\lambda = D_0 \sqrt{ 8 N \epsilon^5 }
$
and not separately on $\epsilon$ or $N$.
For example, the nearest 
neighbour levels spacing distribution $P(s)$ will approach 
$e^{-s}$ when $\lambda \ll 1$.

We have tested this prediction by numerically computing the 
level spacing distribution for different values of 
$\epsilon$ and $N$. One example is shown in Fig.5 for 
which $N \gg N_p$ but since $\lambda \ll 1$  the 
distribution
$P(s)$ is close to $e^{-s}$ as expected. Similar behaviour 
is expected for other quantities such as the two points 
correlation function, the probability distribution of 
eigenfunctions, etc.
The numerical computations are based on the improved 
plane wave decomposition method\cite{Hel}.
The accuracy of eigenvalues is better than one percent of
the mean level spacing. We also compared the results with
the semiclassical formula in order to check that there are no
missing levels.

Notice that the effect predicted here is entirely 
due to quantum dynamical localization and bears no 
relation with the existence of bouncing ball orbits. 
The same behaviour will be present in 
chaotic billiards in which no 
family of periodic orbits exists.

The effects of quantum localization discussed here should be 
observable in microwave or sound wave experiment. 
Finally we would like to mention that the diffusive process 
in angular momentum and the corresponding suppression 
caused by quantum mechanics may 
be 
of interest for a new class of optical 
resonators which have been recently proposed \cite{DS}.

We are indebted to R.Artuso, 
B.V.Chirikov,
I.Guarneri and D.Shepelyansky 
for valuable discussions and suggestions.
B. Li is grateful to the colleagues of the University of Milano
at Como for their hospitality during his visit. His work is
supported in part by the Research Grant Council RGC/96-97/10 and 
the Hong Kong Baptist University Faculty Research Grant
FRG/95-96/II-09 and FRG/95-96/II-92.

\begin{figure}
*\centerline{\epsfxsize 7cm \epsffile{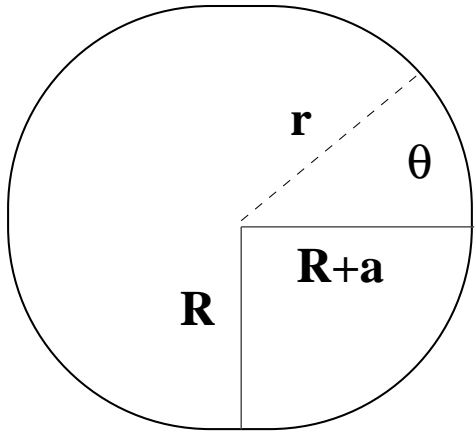}}
\caption{The Bunimovich stadium with radius $R$ and 
straight segments 
$2 a$ ; the variables $(r(\theta), \theta)$ indicate the position
of the point along the boundary.}
\end{figure}

\begin{figure}
*\centerline{\epsfxsize 7cm \epsffile{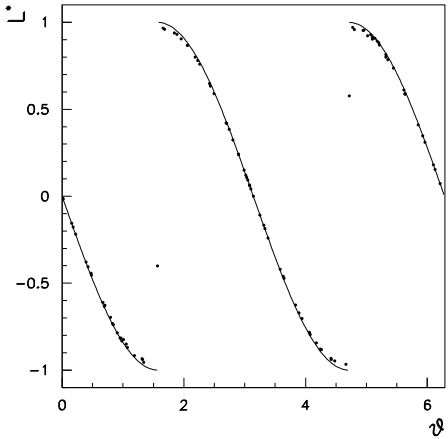}}
\caption{Comparison between the 
billiard dynamics and the map $(6)$. 
Here we plot the variable 
$L^*$ versus $\theta$ (see text).
Points are obtained 
from numerically integrating the motion of 
one particle in the billiard 
for $100$ iterations, starting from $L_0 =0$ and a random
position along the boundary, while the full line
is the function $f(\theta)$ (see text). Here 
$\epsilon=0.01$. The  points not belonging to the curve
are due to collisions with 
one of the straight lines; this occurrence is outside the 
approximation of map $(6)$.}
\end{figure}

\begin{figure}
*\centerline{\epsfxsize 7cm \epsffile{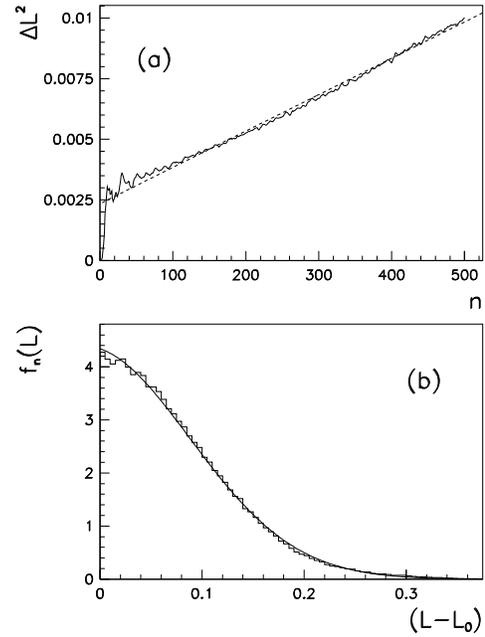}}
\caption{ Diffusion in angular momentum for the
billiard with $\epsilon = 0.01$. Here an
ensemble of $10^4 $ particles was chosen with
initial $L_0 = 0$ and
random position along the boundary.
a) $\Delta L^2 $ as a function of the number of
collisions  $n$;  the dashed line
is the best fit and gives
 $D = \Delta L^2 /n = 1.5 \cdot \quad 10^{-5}$.
b) Distribution function after  $n=500$ collisions
averaged over the last 50 collisions.
The full line is the
best fitting Gaussian with average $-0.016$ and  variance $0.1$.}
\end{figure}

\begin{figure}
*\centerline{\epsfxsize 7cm \epsffile{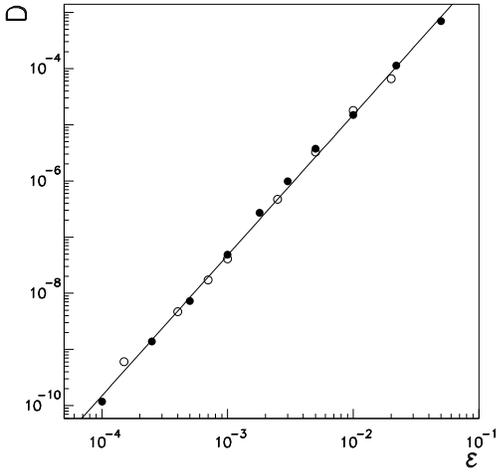}}
\caption{Diffusion coefficient 
$D = \Delta L^2/n$  for the stadium  (full circles)
as a function of $\epsilon$.
Open circles indicate the diffusion rate obtained from the map 
(6).
The line is obtained by the 
usual best fitting procedure
to the true dynamics (full circles) and  gives $D=D_0 \epsilon^{2.5}$
with $D_0= 1.5$.}
\end{figure}

\begin{figure}
*\centerline{\epsfxsize 7cm \epsffile{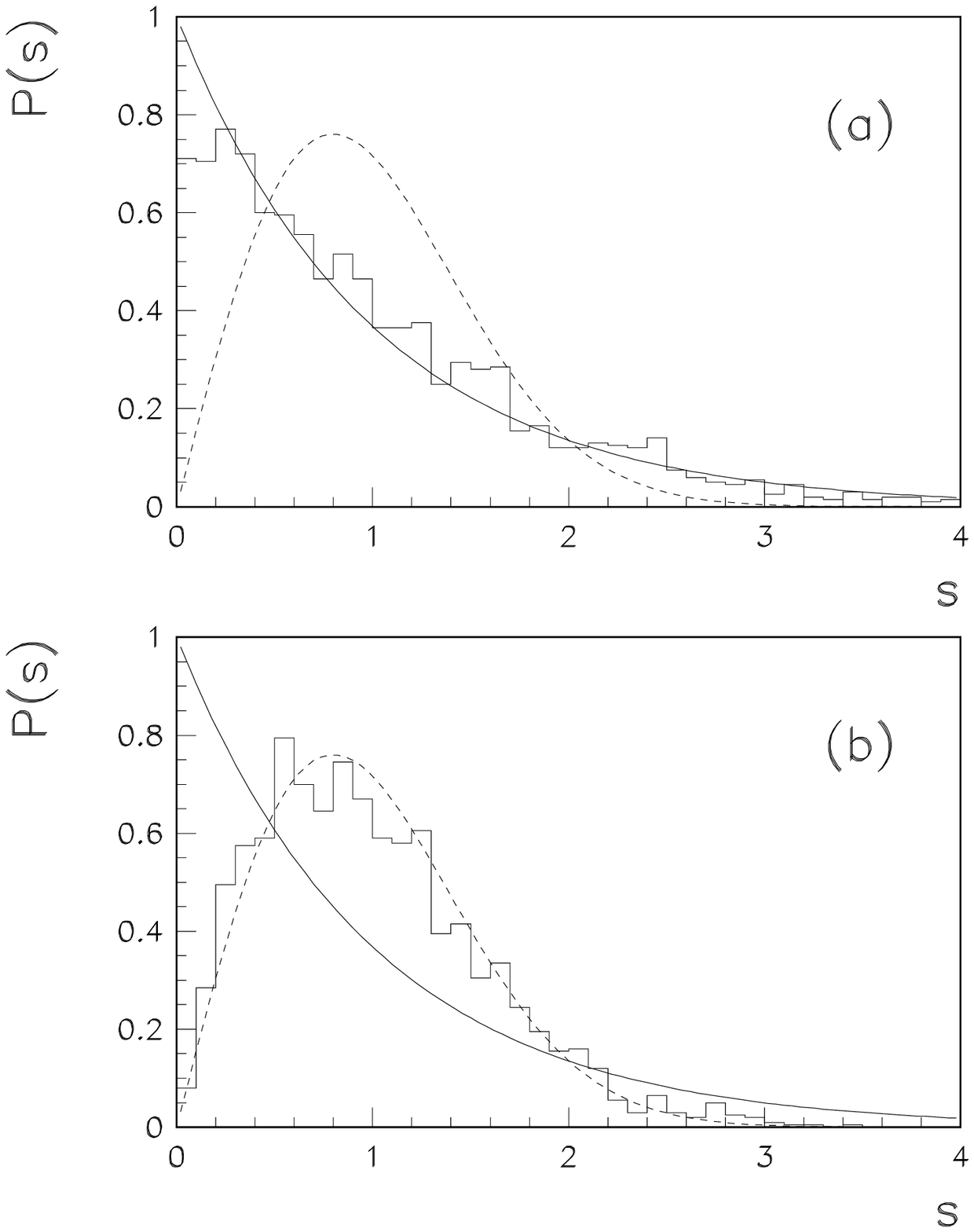}}
\caption{Level spacings distribution 
computed on $2000$ levels in the interval 
$51000 < N < 53000$ for $\epsilon =0.01$ (a) 
and  $\epsilon = 0.1$ (b).
In the first case (a) $N_p \simeq 600$ and $N_{erg} \simeq 2.8 
\cdot 10^8$ and therefore $N_p \ll N \ll N_{erg}$.
The value $\lambda \simeq 0.01$ of the ergodicity parameter 
accounts for the fact that the numerically computed $P(s)$ 
is close to $e^{-s}$ (full curve). 
In the  case (b) one has $N_{erg} \simeq 2.8 \cdot 10^3 \ll N $ 
and therefore, as expected, the distribution $P(s)$ is
close to Wigner-Dyson (dotted curve).
}
\end{figure}


\begin{references}
\bibitem{nota}{It may be interesting to remark that the diffusion 
coefficient computed in terms of the number of collisions, appears to depend
on the initial value of angular momentum $L$. This is due to the fact that
the mean free path  depends on angular momentum and that even 
though the system is ergodic, ergodicity is not uniform in time. If one
computes the diffusion coefficient in terms of real physical time then the 
dependence on the initial $L_0$ value disappears.
For the above reasons map (6) approximate the real dynamics provided 
$L_0$ is not too close to $1$.
}
\end{references}
\end{document}